# Many-body electronic structure in pyrochlore superconductor $CsBi_2$ and spin liquid $Pr_2Ir_2O_7$

**Authors:** Wei Song[1]*, Guowei Liu[2]*, Hanbin Deng[1]†, Tianyu Yang[1], Yongkai Li[3,4,5], Xiao-Yu Yan[1], Ruoxing Liao[1], Qianming Wang[1], Jiayu Xu[2], Chao Yan[1], Yuanyuan Zhao[2], Hailang Qin[2], Da Wang[6], Wenchuan Jing[7,8], Dawei Shen[9], Kosuke Nakayama[10], Takafumi Sato[10], Chandan Setty[11,12], Desheng Wu[2], Boqing Song[13], Tianping Ying[13], Zhaoming Tian[14,15], Akito Sakai[16,17], Satoru Nakatsuji[16,17,18], Harish Kumar[19], Christine A. Kuntscher[19], Zhiwei Wang[3,4,5], Qi-Kun Xue[1,2], Jia-Xin Yin[1,2]†

**Affiliations:**
[1]State Key Laboratory of Quantum Functional Materials, Department of Physics, and Guangdong Basic Research Center of Excellence for Quantum Science, Southern University of Science and Technology, Shenzhen 518055 China.
[2]Quantum Science Center of Guangdong-Hong Kong-Macao Greater Bay Area, Shenzhen, China.
[3]Centre for Quantum Physics, Key Laboratory of Advanced Optoelectronic Quantum Architecture and Measurement (MOE), School of Physics, Beijing Institute of Technology, Beijing 100081, China.
[4]Beijing Key Lab of Nanophotonics and Ultrafine Optoelectronic Systems, Beijing Institute of Technology, Beijing 100081, China.
[5]Material Science Center, Yangtze Delta Region Academy of Beijing Institute of Technology, Jiaxing, 314011, China.
[6]National Laboratory of Solid State Microstructures & School of Physics, Nanjing University, Nanjing 210093, China.
[7]State Key Laboratory of Functional Materials for Informatics, Shanghai Institute of Microsystem and Information Technology, Chinese Academy of Sciences, Shanghai 200050, China.
[8]Shanghai Synchrotron Radiation Facility, Shanghai Advanced Research Institute, Chinese Academy of Sciences, Shanghai 201210, China.
[9]National Synchrotron Radiation Laboratory, University of Science and Technology of China, 42 South Hezuohua Road, Hefei, Anhui 230029, China.
[10]Advanced Institute for Materials Research (WPI-AIMR), Tohoku University, Sendai 980-8577, Japan.
[11]Department of Physics and Astronomy, Iowa State University, Ames, Iowa 50011, USA.
[12]Ames National Laboratory, U.S. Department of Energy, Ames, Iowa 50011, USA.
[13]Beijing National Laboratory for Condensed Matter Physics, Institute of Physics, Chinese Academy of Sciences, Beijing, China.
[14]School of Physics and Wuhan National High Magnetic Field Center, Huazhong University of Science and Technology, Wuhan, Hubei 430074, China.
[15]Shenzhen Huazhong University of Science and Technology Research Institute, Shenzhen 518057, China.
[16]The Institute for Solid State Physics, The University of Tokyo, Kashiwa, Chiba 277-8581, Japan.
[17]Trans-Scale Quantum Science Institute, The University of Tokyo, Bunkyo, Tokyo 113-0033, Japan.
[18]Department of Physics, The University of Tokyo, Bunkyo, Tokyo 113W-0033, Japan.
[19]Experimentalphysik II, Institute for Physics, University of Augsburg, 86159 Augsburg, Germany
*These authors contributed equally to this work.
†Corresponding authors. E-mail: denghb@sustech.edu.cn; yinjx@sustech.edu.cn

**The pyrochlore lattice materials can exhibit geometrical frustration, while the related many-body electronic states remain elusive. In this work, we performed scanning tunneling microscopy measurements on the pyrochlore superconductor $CsBi_2$ and spin liquid $Pr_2Ir_2O_7$ at 0.3 K. For the first time, we obtained atomically resolved images of their (111) surfaces, revealing a hexagonal lattice or a kagome lattice. Tunneling spectroscopy in $CsBi_2$ reveals a nearly fully opened superconductivity gap. The ratio of $2\Delta/k_BT_C = 4.7$ suggests relatively strong coupling superconductivity, as compared with that in kagome superconductors $AV_3Sb_5$ (A = K, Rb, Cs). In contrast to the previous study categorizing $CsBi_2$ as a type-I superconductor, the applied magnetic field induces a hexagonal vortex lattice in which each vortex core exhibits an intriguing three-fold symmetry state. In $Pr_2Ir_2O_7$, we observed a spatially homogeneous Kondo-lattice resonance, which is compared with that in the kagome Kondo-lattice material $CsCr_6Sb_6$. We further discover that the Kondo resonance exhibits a spatial modulation with three-fold symmetry, and the applied magnetic field induces a Zeeman splitting of the Kondo resonance with intriguing atomic site dependence. We discuss the relations of these many-body electronic phenomena with the pyrochlore lattice geometry and its charge or spin frustration. Our systematic observations offer atomic-scale insights into the many-body electronic structures of the geometrically frustrated pyrochlore superconductors and spin liquids.**

**I Introduction**

A pyrochlore lattice is made of corner-sharing tetrahedra [Fig. 1(a)] and serves as the three-dimensional analog of the kagome lattice made of corner-sharing triangles. Quantum materials hosting a pyrochlore lattice exhibit various correlated phases [1-23], including unconventional superconductivity, non-Fermi liquid behaviors, spin ices states, topological fermions, and flat bands. The pyrochlore quantum materials can exhibit both spin and charge frustrations. The local spins in the pyrochlore lattice can form spin ice [Fig. 1(a)], where the spins follow two-in-two-out ice rules, analogous to the behavior of electric dipoles in water ice featuring multifold degeneracy. The destructive interferences of the electron wave function from its sublattice lead to the flat band with quenched kinetic energy, as demonstrated in its tight-binding model considering nearest electron hopping [Fig. 1(b)]. Historically, the earliest superconducting pyrochlore materials, such as $CeRu_2$, and its related compounds, were initially proposed as ferromagnetic superconductors [1,2] by Matthias, Suhl, and Anderson in the 1950s. These findings highlighted their unconventional superconducting properties and initiated extensive research into superconductivity coexisting with magnetism. In these materials, the localized $4f$ electrons of Ce contribute to flat bands, while the $4d$ electrons of Ru in the pyrochlore lattice can also form geometrical frustration induced flat bands, further complicating the understanding of their unusual flat-band magnetism. In this work, we selected single-crystal $CsBi_2$ [17], which shares a crystal structure similar to that of $CeRu_2$ but lacks contributions from $f$ electrons. Another striking phenomenon observed in pyrochlore lattices is spin ice magnetism, first identified in $Ho_2Ti_2O_7$ [3] and related compounds that are insulators. In this work, we selected $Pr_2Ir_2O_7$, which shares a similar crystal structure with $Ho_2Ti_2O_7$ but features Ir $5d$ electron mediated metallicity and local Pr $4f$ spins without long-range magnetic order, and it has been identified as a metallic spin liquid [5] with an intriguing spontaneous Hall effect [7]. In this study, we investigate the atomically resolved electronic structures of the pyrochlore superconductor $CsBi_2$ and spin liquid $Pr_2Ir_2O_7$ using high-resolution scanning tunneling microscopy down to 0.3K.

**II Methods**

Our tunneling experiments were performed at a lattice temperature of 280 mK unless otherwise specified. The electronic temperature for high-resolution measurements was estimated at best to be 310 mK based on the measurement of the superconducting gap of Pb, and is further limited by the modulation voltage as specified for each data. Single crystals were cleaved mechanically *in situ* at 10

K in ultra-high vacuum conditions, and then immediately inserted into a microscope head pre-cooled to the He4 base temperature (4.2 K). The microscope was further cooled to 0.3 K using a He3-based single-shot refrigerator. Each cleaved crystal was extensively searched for over a week to identify an atomically flat cleavage surface larger than 1μm×1μm, with more than 40 crystals being cleaved and analyzed. The magnetic field was applied at a controlled ramping speed of 1T every 30 mins. After ramping the field to a desired value, the superconducting magnet is set to the persistent mode, and the system is allowed to relax for 1–2 hours before locating the same atomic position and commencing spectroscopic measurements. Tunneling conductance spectra were obtained with Ir/Pt tips and standard lock-in amplifier techniques. Single crystals of $CsBi_2$ were synthesized using a self-flux method with Bi as the flux. High-purity Bi (Alfa Aesar, 99.999%) and Cs (Alfa Aesar, 99.9%) were loaded into an alumina crucible, then sealed in an evacuated quartz tube. The tube was heated to 650 °C very slowly and maintained at this temperature for 8 hours to ensure thorough homogenization. Subsequently, it was cooled down to 300 °C at a rate of 5 K/h, where the excess Bi flux was removed using a centrifuge. Finally, hexagonal, shiny crystals were obtained, typically measuring approximately 2×2×2 $mm^3$. Single crystals of $Pr_2Ir_2O_7$ were grown with the method in Ref. 23.

**III Realization of atomic surface lattice in pyrochlore lattice materials**

Figure 1(c) illustrates the pyrochlore lattice structure, and its (111) cleavage surface can be either a hexagonal lattice or a kagome lattice. Figure 1(d) depicts the crystal structure of $CsBi_2$, which is classified as a cubic Laves phase wherein Bi atoms form a pyrochlore lattice network. The stacking sequence of $CsBi_2$ along the [111] direction is: $Bi_3$ kagome lattice-Cs hexagonal lattice-Bi hexagonal lattice-Cs hexagonal lattice. Through extensive searches, we identified atomically flat (111) cleavage surfaces, as shown in Fig. 1(e). Individual surface atoms can be resolved, forming a hexagonal lattice with half of the atoms missing. These missing atoms generate stripe-like features that frequently form domain structures. We propose that the cleavage of a single crystal produces two cleavage surfaces, each populated by half of the atoms from the Bi hexagonal layer. This unique distribution of the Bi atoms results in charge-neutral surfaces, as illustrated in Fig. 1(f). Figure 1(g) illustrates the crystal structure of $Pr_2Ir_2O_7$. Ignoring O atoms that are often invisible in the topographic image, two possible terminations exist: $Pr_3Ir$ surface with Pr atoms forming a kagome lattice stuffed with Ir hexagonal lattice, or $PrIr_3$ surface with Ir atoms forming a kagome lattice stuffed with Pr hexagonal lattice. Due to the significantly larger atomic radius of Pr compared to Ir, we expect the former to resemble a kagome lattice, whereas the latter appears as a hexagonal lattice. In our systematic scanning, we either observe a kagome-structured surface, as depicted in Figs. 1(h) and (i), or a highly disordered hexagonal lattice surface similar to that in Ref. 15. We thus empirically identify the kagome surface as the $Pr_3Ir$ layer and focus on this clean surface. Compared with the highly disordered surface obtained through room temperature cleavage [15], the cryogenic cleavage produces an atomically ordered surface suitable for state-of-the-art spectroscopic characterization. After successfully resolving their atomic-scale surface structures for both pyrochlore lattice materials, we further investigated their electronic structures.

**IV Superconducting electronic structure and vortex matter in $CsBi_2$**

The superconducting transition temperature $T_C$ of $CsBi_2$ was determined to be 4.5 K by susceptibility measurements under zero-field cooling conditions [Fig. 2(a)], in agreement with previous findings [17]. The differential conductance spectrum acquired from the (111) surface reveals a U-shaped energy gap, with coherence peaks observed at ±0.95 meV [Fig. 2(b)]. We also show a spectrum of superconducting Pb as a reference for the instrumental resolution of our system. This gap feature progressively vanishes as the temperature is raised above $T_C$ [Fig. 2(b)] or when an external magnetic field is applied perpendicular to the surface [Fig. 2(c)], indicating its direct relationship with superconductivity. We note that the surface reconstruction aligns with the charge-neutral requirement, and the observed signal may reflect bulk superconducting properties. We systematically measured the superconducting gap at various locations with atomic resolution by employing line-cut spectra [Fig. 2(d)] and spectroscopic mapping [Figs. 2(e)], both revealing that the pairing gap is spatially homogeneous. We also point out that although the U-shaped pairing gap strongly deviates from the line-node gap structure, the gap bottom is not perfectly flat. The weak in-gap excitations may reflect a weak anisotropy of the pairing gap in momentum space. Collectively, our observations suggest the existence of a nearly fully opened pairing gap.

We analyze the superconducting coupling strength of $CsBi_2$ and compare it to previously investigated geometrically frustrated kagome superconductors [24-27]. In Fig. 2(f), we plot the energy gap as a function of $T_C$. The energy gap values were extracted from coherence peak-to-peak distances obtained through tunneling spectroscopy. For pyrochlore and kagome superconductors, the tip can occasionally capture a superconducting flake, causing an overestimation of the measured pairing gap. To prevent this artifact, we ensured that our data were unaffected by such issues. We determined that the coupling strength for the pyrochlore superconductor $CsBi_2$ is $2\Delta/k_BT_C = 4.74$, exceeding that in $AV_3Sb_5$ kagome superconductors extracted with similar methods. The existence of flat bands in pyrochlore and kagome lattices can potentially enhance the electron correlation, leading to superconductivity with a stronger coupling strength [28,29]. In $AV_3Sb_5$ superconductors, the flat band exists near 1 eV well away from the Fermi level [30], while recent angle-resolved photoemission finds that the nearest flat band in $CsBi_2$ is near -0.1 eV [22], which can be one reason for its stronger coupling strength.

We perturb the pairing state with a magnetic field of 0.05T and perform spectroscopic imaging of the zero-energy in-gap state over a large field of view (1μm×1μm), a task that is particularly challenging for three-dimensional materials. Contrary to previous claims categorizing $CsBi_2$ as a type-I superconductor [17], our results unambiguously show that an applied magnetic field induces vortices that form a hexagonal vortex lattice, as shown in Fig. 3(a). From this spectroscopic image, we observed that each vortex core state exhibits anisotropy. To investigate this anisotropy clearly, we apply a small magnetic field of 0.01T to minimize overlap between vortex states. Under this reduced field, the spectroscopic image of an isolated vortex state [Fig. 3(b)] reveals an unusual three-fold symmetry. Systematic differential conductance measurements across the vortex core revealed no apparent sharp bound states within the pairing gap [Figs. 3(c) and (d)]. Instead, the in-gap state is nearly uniformly elevated at the vortex core center, suggesting that the superconductor resides outside the quantum clean limit. In this dirty limit regime, the mean free path exceeds the superconducting coherence length, and the lifetime broadening is larger than the superconducting energy gap.

We further analyzed the decay of the vortex core state in Fig. 3(e) by fitting the data with an exponential decay function A·exp(-$r/\xi$)+B, where A and B are fitting parameters and $\xi$ is the superconducting coherence length. The extracted superconducting coherence lengths along the *a*-axis and *-a*-axis are 64nm and 90nm, respectively. We also calculated the upper critical field $H_{C2}$ using the relation $H_{C2} = \frac{\Phi_0}{2\pi\xi^2}$, where $\Phi_0$=2.07×10$^{-7}$ G cm$^2$ is the flux quantum. Based on these coherence lengths, the calculated $H_{C2}$ values are 80mT and 41mT, respectively. They align with our experimental observation of the superconducting gap feature disappearing at B=60mT [Fig. 2(c)], highlighting the internal consistency of our measurement. The three-fold anisotropy observed in the vortex core state implies a corresponding three-fold symmetry in the coherence length. The coherence length $\xi$ is proportional to both Fermi velocity $v_F$ and inversely proportional to the energy gap $\Delta$, given by $\xi = \frac{\hbar v_F}{\pi\Delta}$. Since the U-shaped gap in Fig. 2 indicates a relatively isotropic energy gap, the anisotropy observed is likely attributable primarily to variations in the Fermi velocity. For this multiband system, the largest Fermi surface plays a crucial role in determining the vortex geometry. We plotted the largest Fermi surface in Fig. 3(f), color-coded according to the Fermi velocity obtained from the first-principles calculations. Notably, the Fermi velocity exhibits pronounced three-fold symmetry, consistent with our observations of the anisotropic in the vortex core state. The three-fold symmetry of the Fermi velocity in three dimensions is fundamentally related to the three-dimensional pyrochlore lattice geometry.

**V Homogeneous Kondo-lattice resonance and magnetic field response in $Pr_2Ir_2O_7$**

We then investigated the electronic structure of $Pr_2Ir_2O_7$, in which coupling between localized Pr 4$f$ moments and itinerant Ir 5$d$ electrons results in Kondo-lattice behavior characterized by a resistivity upturn near 40 K [5]. On the atomically resolved kagome surface, we observed a prominent electronic state located slightly below the Fermi level and a dip feature at the Fermi level, as shown in Fig. 4(a). These spectroscopic features appear spatially homogeneous on the atomically resolved surface that we attribute to the intrinsic Kondo-lattice behavior [31-39], in contrast to the strongly inhomogeneous tunneling spectra previously observed on disordered surfaces in $Pr_2Ir_2O_7$ [15]. Furthermore, the resonance intensity shows slight modulation correlated with the lattice periodicity. Spectroscopic imaging at the resonance peak energy reveals clear three-fold symmetry concerning the kagome lattice center, as shown in the simultaneously obtained data in Figs. 4(b) and (c). Although the kagome lattice itself exhibits six-fold symmetry, the underlying pyrochlore lattice at the kagome center possesses intrinsic $C_{3V}$ symmetry, as illustrated in the inset of Fig. 4(c), which likely accounts for the observed three-fold symmetry.

The spectral asymmetry of the resonance prompted us to employ the Fano equation $F(E) \approx \frac{(q+E/\Gamma)^2}{1+(\frac{E}{\Gamma})^2}$, where $q$ is the dimensionless parameter characterizing the coupling strength between the tip and discrete states, and $\Gamma$ denotes the resonance width. The spatially averaged dI/dV spectral peak can be well-fitted by $A*F(E-E_0)+B$, as shown in Fig. 4(d), with A, B, and $E_0$ as additional adjustment parameters. We note that the dip is not captured by this fitting, which may require a theoretical model with extra parameters [40,41]. The obtained resonance width $\Gamma$ = 4.5 meV defines the characteristic energy scale of the resonance. By associating this state with a Kondo-lattice resonance, we estimated the spectroscopic Kondo temperature as $T_K = \Gamma/1.4k_B$ = 37 K. Our temperature-dependent measurements show that the resonance peak progressively broadens and weakens as temperature increases [Fig. 4(e), solid lines]. In tunneling experiments, the dI/dV spectra measure the convolution of the density of states and the derivative of the Fermi-Dirac distribution function. As such, we also plot the temperature convolution of the spectrum taken at our lowest temperature (T = 0.3 K) for each temperature (dashed lines) for comparison. The experimental data exhibit stronger temperature-induced broadening than the convoluted curves, indicating that the resonance is interaction-driven and intrinsically temperature-dependent. The resonance disappears near 40 K, consistent with the previously reported resistivity upturn temperature [5] or optical conductivity maximum [23], and the critical temperature derived from our Fano fitting. We have noticed that the peak's energy shifts to the higher binding energies with increasing temperature, which we attribute to the asymmetry of the peak profile as similar shift trend to higher energies is produced by the thermal convolution simulation (dashed lines).

We also compare its spectrum with that in the recently discovered kagome Kondo-lattice material $CsCr_6Sb_6$ without magnetic order [42]. Similar to the cleaving of $CsV_3Sb_5$, the cleavage of $CsCr_6Sb_6$ can produce Cs hexagonal lattice surface and Sb honeycomb lattice. We focus on the Sb honeycomb lattice that is tightly bonded with the underlying bilayer Cr kagome lattice, as shown in Fig. 4(f). The tunneling spectrum of $CsCr_6Sb_6$ exhibits a peak just below the Fermi level and a strong dip at the Fermi level [Fig. 4(g)]. These observations align with our expectation for the Kondo-lattice behavior in $CsCr_6Sb_6$. We have also measured its related material $CsCr_3Sb_5$ that is magnetically ordered [43], and we do not detect such Kondo-lattice behavior. Such features are spatially homogeneous with slight modulations [Fig. 4(g)], and these spectroscopic features disappear with elevating the temperature to 60 K~70 K [Fig. 4(h)], roughly consistent with the resistivity upturn temperature. We compare the spectrums between $CsCr_6Sb_6$ and $Pr_2Ir_2O_7$. Although occurring at a similar energy, its resonance peak is less pronounced than that in $Pr_2Ir_2O_7$. The resonance peak (many-body flat band) in Kondo-lattice is often understood as the interplay between single-particle flat band and itinerate electrons under correlation. The stronger resonance in $Pr_2Ir_2O_7$ may be related with the strongly localized $f$ electrons in $Pr_2Ir_2O_7$, as compared with the geometrical frustration induced flat band in $CsCr_6Sb_6$. On the other hand, the gap is deeper than that in $Pr_2Ir_2O_7$. Such gap can often be understood as a hybridization gap between the many-body flat band and dispersive bands. In $CsCr_6Sb_6$ both the flat band and itinerate electrons are from Cr 3$d$ electrons, which may be the reason for a stronger hybridization gap, as compared with hybridization with different orbitals in $Pr_2Ir_2O_7$. Although further rigorous theoretical modeling is needed, these comparisons provide a heuristic experimental reference for studying Kondo-lattice behavior in geometrically frustrated systems.

Lastly, taking advantage of the atomically resolved Kondo-lattice resonance, we examine the magnetic field dependence of the resonance at three distinct atomic sites in $Pr_2Ir_2O_7$ on the kagome lattice, as illustrated in Fig. 5(a). The magnetic field was applied perpendicular to the surface and along the crystalline (111) direction. In previous tunneling experiments on other Kondo-lattice candidates [35-38], where a Kondo-lattice resonance exhibiting a Fano line shape was observed, no clear splitting of the resonance under strong magnetic fields was detected, although broadening occurred. Thus the Kondo lattice generically features an effective g factor g<<1 in experiments. Here, we also find no detectable splitting up to 8T at both sites A and B, as shown in Fig. 5(b) and (c), respectively. However, we observe a progressive splitting of the resonance at the center of the kagome pattern (site C), as shown in Fig. 5(d). When we reversed the direction of the applied magnetic field, we observed a similar resonance splitting at site C, as shown in Fig. 5(e). The splitting is more clearly revealed through derivative analysis of the magnified spectrum shown in Fig. 5(f). The splitting at the C site at 8T is approximately 1.6 meV, corresponding to an effective $g$-factor of about 1.8, which is enhanced from g<<1 at the generical Kondo lattice sites.

We discuss possible origins of the site-dependent magnetic-field splitting observed in its Kondo-lattice resonance. The minimum field with which the resonant state splits is related to the Kondo temperature by approximately $g\mu_B B \approx 0.5 k_B T_K$ [44,45]. For an estimated $T_K \sim 40$ K and g ~ 2, a minimum field of approximately 15T would be required. This estimation aligns with our observation at sites A and B, but not C, especially noting that these spectrums at these locations have similar resonance peak width, thus similar estimated $T_K$. In our surface identification (Fig. 1), the observed kagome lattice (sites A and B) consists of Pr atoms possessing local moments, whereas the central site C corresponds to Ir atoms. Note that people have not detected a site-dependent Zeeman effect in other Kondo-lattice systems, so we think the chemical difference may not be a key reason, and we discuss the potential reason based on the unique geometry of the current system. The unique geometry of the pyrochlore or kagome lattice leads to significant frustration of the Pr local moments, resulting in spin-ice formation without long-range order [7-11]. The coupling of such frustrated magnetism with itinerate electrons as metallic spin liquid is unprecedented as compared with previous Kondo-lattice study with scanning tunneling microscopy under magnetic fields [35-38]. Particularly, a giant anomalous Hall effect has been detected in $Pr_2Ir_2O_7$, which is attributed to the spin chirality effect on the Ir sites from the unusual spin texture of Pr 4$f$ moments [7]. Such a quantum state is proposed to be chiral spin liquid as a result of a melted spin ice [7]. Therefore, there can be an internal magnetic field at the Ir sites from the unusual spin texture of the Pr 4$f$ moments that facilitates its Zeeman splitting behavior under a relatively smaller external magnetic field as compared with sites A and B or other Kondo-lattice systems. This geometrical Zeeman effect deserves further theoretical investigation based on the framework of chiral spin liquid.

**Summary**

In summary, for both $CsBi_2$ and $Pr_2Ir_2O_7$, we successfully overcome substantial experimental challenges to achieve atomically resolved cleaving surfaces. The electronic structures of both materials exhibited clear three-fold symmetry, which we attribute to the effect of the underlying pyrochlore lattice symmetry on surface. We demonstrated that $CsBi_2$ behaves as a type-II superconductor, evidenced by vortex formation under applied magnetic fields, and that $Pr_2Ir_2O_7$ exhibits a spatially homogeneous Kondo-lattice resonance on atomically cleaner surface lattices. We further identified strong-coupling superconductivity and site-dependent Zeeman splitting of the Kondo-lattice resonance. We discussed the potential link between these intriguing phenomena and the intrinsic geometrical frustration of the pyrochlore lattice. This progress encourage us to further explore quantum entanglement with scanning tunneling microscopy in frustrated quantum materials in the future.

**Figures and captions**

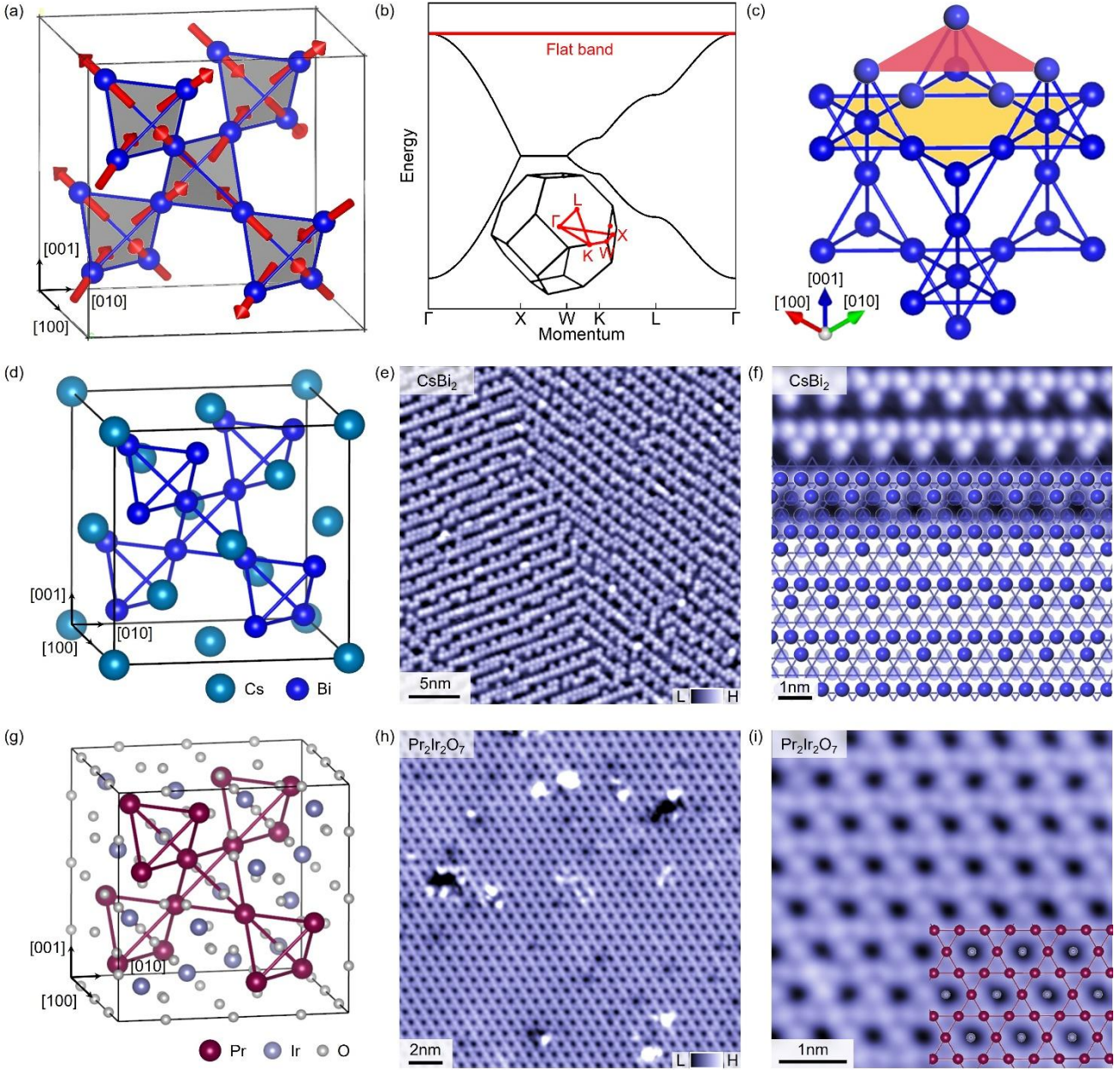

Figure 1 Atomically resolved surfaces of pyrochlore superconductor CsBi$_2$ and spin liquid Pr$_2$Ir$_2$O$_7$. (a) Crystal structure of a pyrochlore lattice. Red arrows indicate spins forming spin-ice magnetism on the pyrochlore lattice. (b) Tight-binding model of a pyrochlore lattice considering nearest-neighbor hopping interactions, showing a characteristic flat band. The inset identifies high-symmetry points within the Brillouin zone. (c) Schematic illustrating that the (111) surface of a pyrochlore lattice can terminate as either a hexagonal lattice (red plane) or a kagome lattice (yellow plane). (d) Crystal structure of CsBi$_2$. (e) Topographic image showing the atomically resolved (111) cleavage surface of CsBi$_2$, exhibiting stripe-like features with clearly defined domain boundary. We use bias voltage $V = 50$ mV and tunnelling current $I = 500$ pA. (f) Structural assignment of the surface atoms to the CsBi$_2$ crystal structure, highlighting the absence of half of the Bi atoms in the hexagonal layer (marked by transparent spheres), creating a charge-balanced surface. Blue lines delineate the underlying kagome lattice-structure. We use $V = 50$ mV and $I = 500$ pA. (g) Crystal structure of Pr$_2$Ir$_2$O$_7$. (h) Topographic image showing the atomically resolved (111) cleavage surface of Pr$_2$Ir$_2$O$_7$. We use $V = 100$ mV and $I = 500$ pA. (i) Magnified view of the topography shown in (h), clearly resolving the kagome lattice. We use $V = 100$ mV and $I = 500$ pA. All data are taken at 0.3 K.

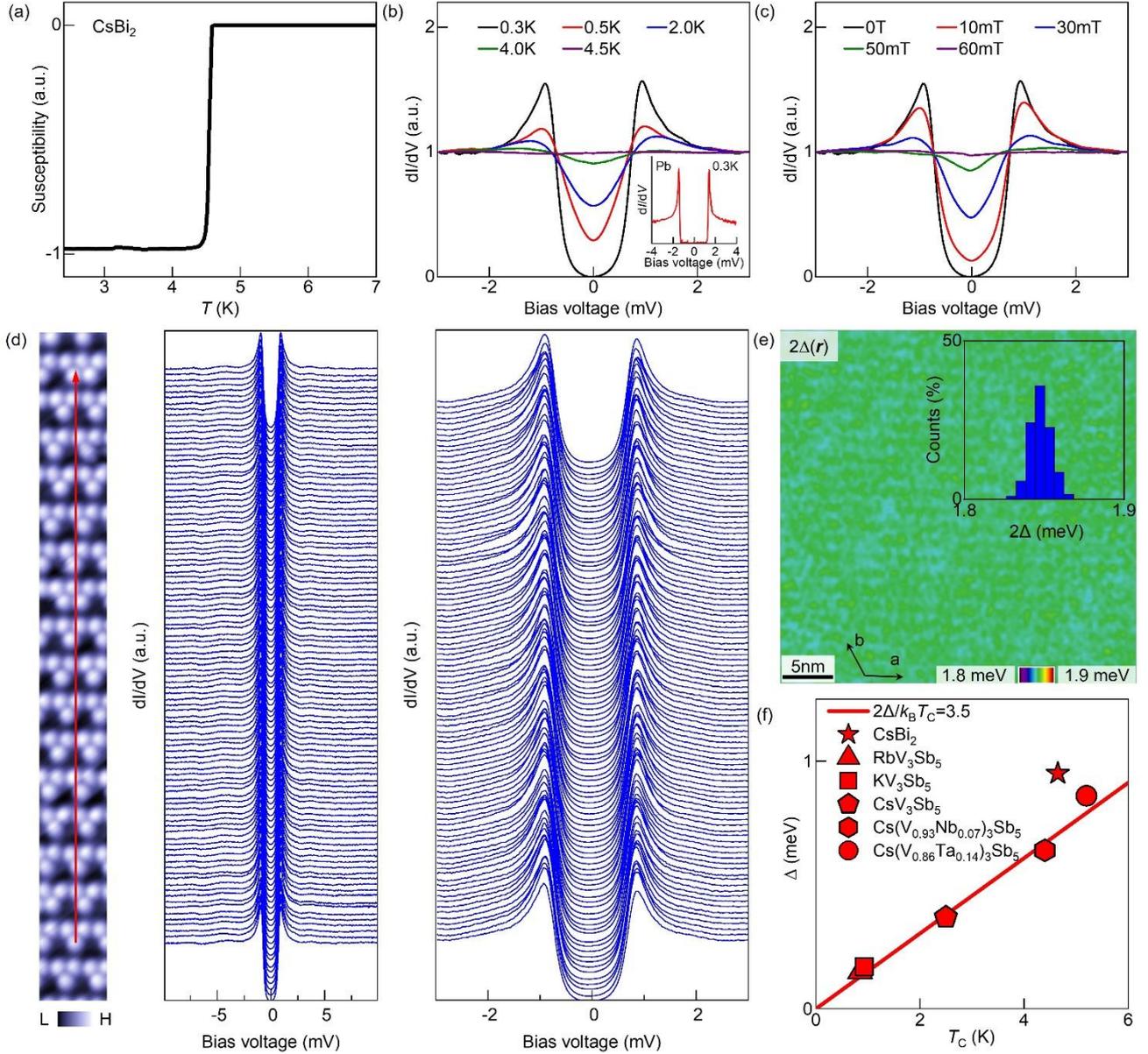

Figure 2 Full opened superconducting gap and strong-coupling superconductivity in CsBi$_2$. (a) Susceptibility measured under zero-field-cooling conditions, showing superconducting a superconducting transition at 4.5 K. (b) Differential conductance spectra illustrating the temperature dependent evolution of the superconducting gap. We use $V = 3$ mV, $I = 1$ nA and modulation voltage $V_m = 0.1$ mV. The inset shows a spectrum of Pb sample measured at the same instrument. (c) Differential conductance spectra demonstrating the suppression of the superconducting gap with increasing magnetic field. Spectra were acquired between vortices. We use $V = 3$ mV, $I = 1$ nA, $V_m = 0.1$ mV. (d) Differential conductance spectra recorded along the indicated line in the topographic image (left). The middle panel displays spectra at a broader energy scale ($V = 30$ mV, $I = 1$ nA, $V_m = 0.3$ mV), while the right panel shows spectra at a narrower energy scale ($V = 3$ mV, $I = 1$ nA, $V_m = 0.1$ mV). Spectra are vertically offset for clarity. (e) Spatial mapping of the superconducting gap map $\Delta$ over the same region. The inset presents the statistics distribution of measured gap values. We use $V = 3$ mV, $I = 0.5$ nA and $V_m = 0.1$ mV. (f) Superconducting gap $\Delta$ plotted against $T_C$, compared with kagome superconductors, emphasizing the strong-coupling superconductivity observed in CsBi$_2$. All data are taken at 0.3 K.

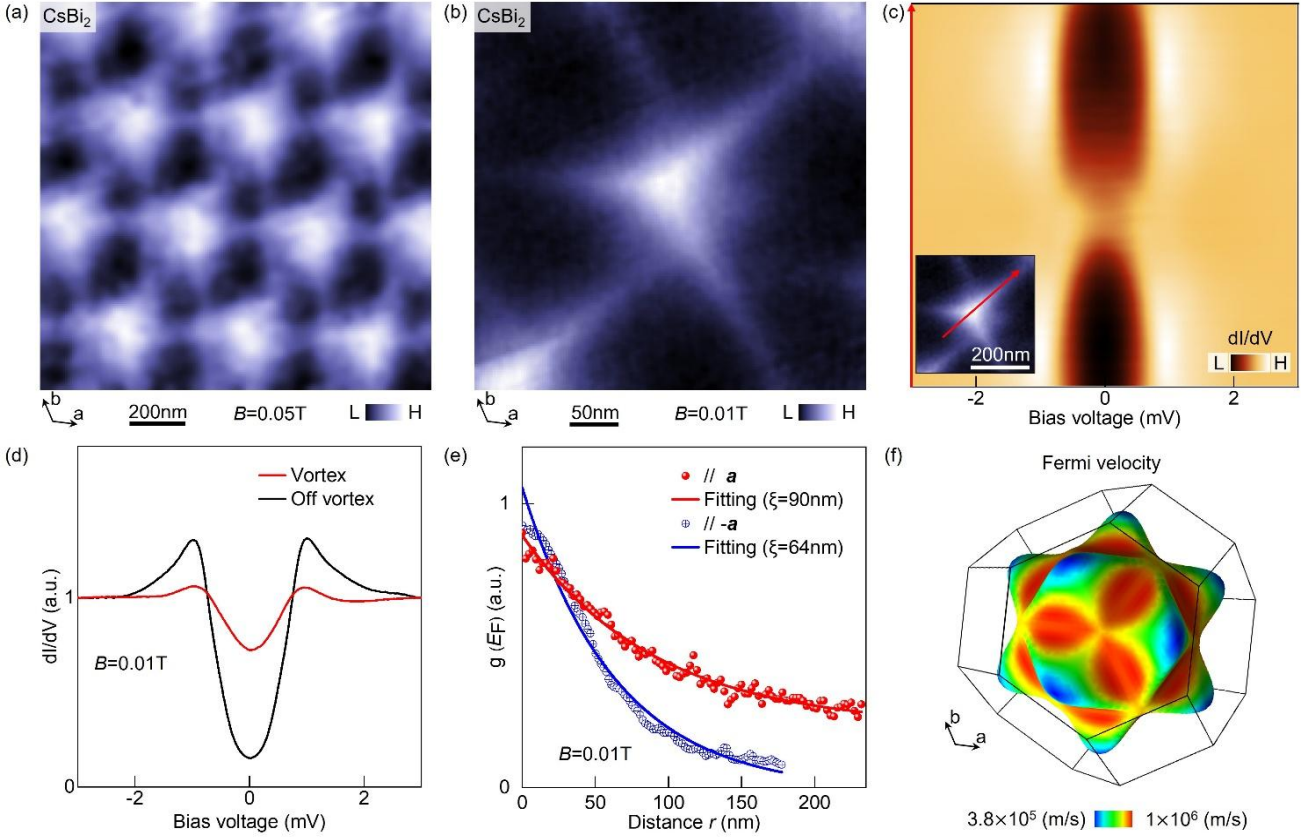

Figure 3 Hexagonal vortex lattice and anisotropic vortex core states in CsBi$_2$. (a) Zero-energy differential conductance map acquired over a large field of view (1μm×1μm) at an applied perpendicular magnetic field of B=0.05T, revealing a hexagonal vortex lattice. We use $V = 10$ mV, $I = 100$ pA, and $V_m = 0.3$ mV. (b) Zero-energy differential conductance map measured at a lower magnetic field (B = 0.01T), revealing an isolated vortex core state exhibiting distinct three-fold symmetry. We use $V = 10$ mV, $I = 100$ pA, and $V_m = 0.3$ mV. (c) Intensity plot of the differential conductance spectra along the line indicated in the inset, traversing a single vortex core. We use $V = 3$ mV, $I = 1$ nA, and $V_m = 0.1$ mV. (d) Representative differential conductance spectra measured at the vortex core center and away from the core. We use $V = 3$ mV, $I = 1$ nA, and $V_m = 0.1$ mV. (e) Spatial decay profile of a single vortex core state along *a* axis (red data) and -*a* axis (blue data). Solid lines represent exponential fits used to estimate the coherence lengths. (f) First-principles calculation of the largest Fermi surface in CsBi$_2$, colored according to the calculated Fermi velocity. Smaller Fermi surfaces have been omitted for clarity. The Fermi velocity shows pronounced three-fold symmetry along the (111) direction. All data are taken at 0.3 K.

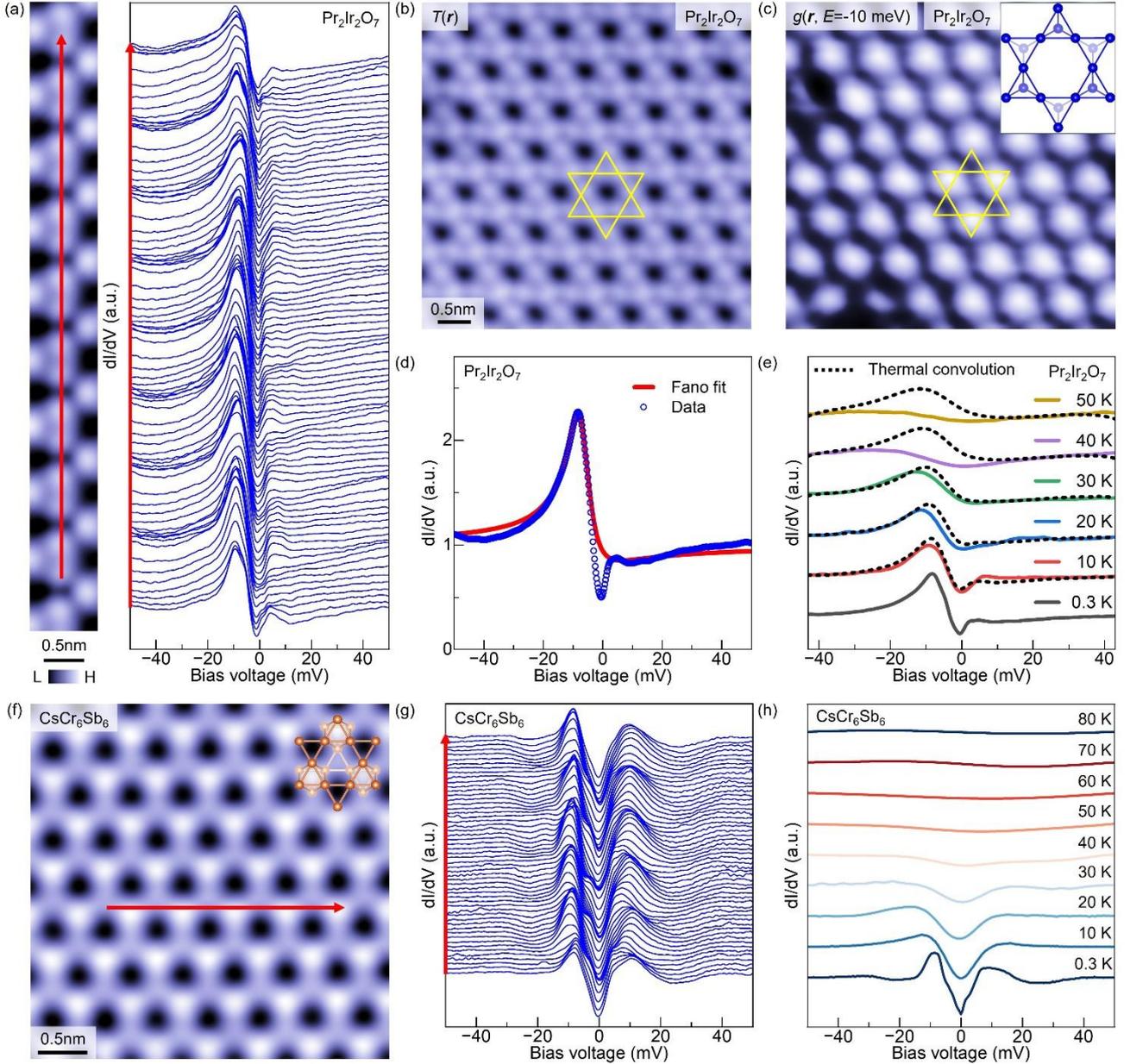

Figure 4 (a) Differential conductance spectra recorded along the indicated line in the atomically resolved kagome lattice shown at left. Spectrums are offset for clarity. $V = 50$ mV, $I = 500$ pA, and $V_m = 0.3$ mV. (b) Topographic image of the atomically resolved kagome lattice surface. $V = 50$ mV, $I = 500$ pA. (c) Differential conductance mapping acquired at the resonance peak energy, revealing clear three-fold symmetry with respect to the kagome lattice center. $V = 50$ mV, $I = 500$ pA, and $V_m = 0.3$ mV. Inset illustrates the atomic configuration of the lattice and underlying pyrochlore lattice atoms, highlighting the origin of emergent three-fold symmetry. (d) Spatially averaged differential conductance spectrum fitted using a Fano resonance line shape. (e) Temperature-dependent differential conductance spectra (solid lines). Dashed lines represent spectra simulated by thermally convoluting the lowest-temperature spectrum. Spectra are offset for clarity. $V = 50$ mV, $I = 100$ pA, and $V_m = 0.3$ mV. (f) Topographic image of the Sb honeycomb lattice in $CsCr_6Sb_6$, tightly coupled with the underlying . Cr kagome lattice. Inset (upper right) illustrate the underlying Cr kagome bilayers. $V = -50$ mV, and $I = 500$ pA. (g) Differential conductance spectra recorded along the red line indicated (f). Spectra are offset for clarity. $V = 50$ mV, $I = 1$ nA, and $V_m = 0.5$ mV. (h) Temperature dependent differential conductance spectra. Spectrums are offset for clarity. $V = -50$ mV, $I = 1$ nA, and $V_m = 1$ mV. All data are taken at 0.3 K, except for (g) and (h) with more temperatures.

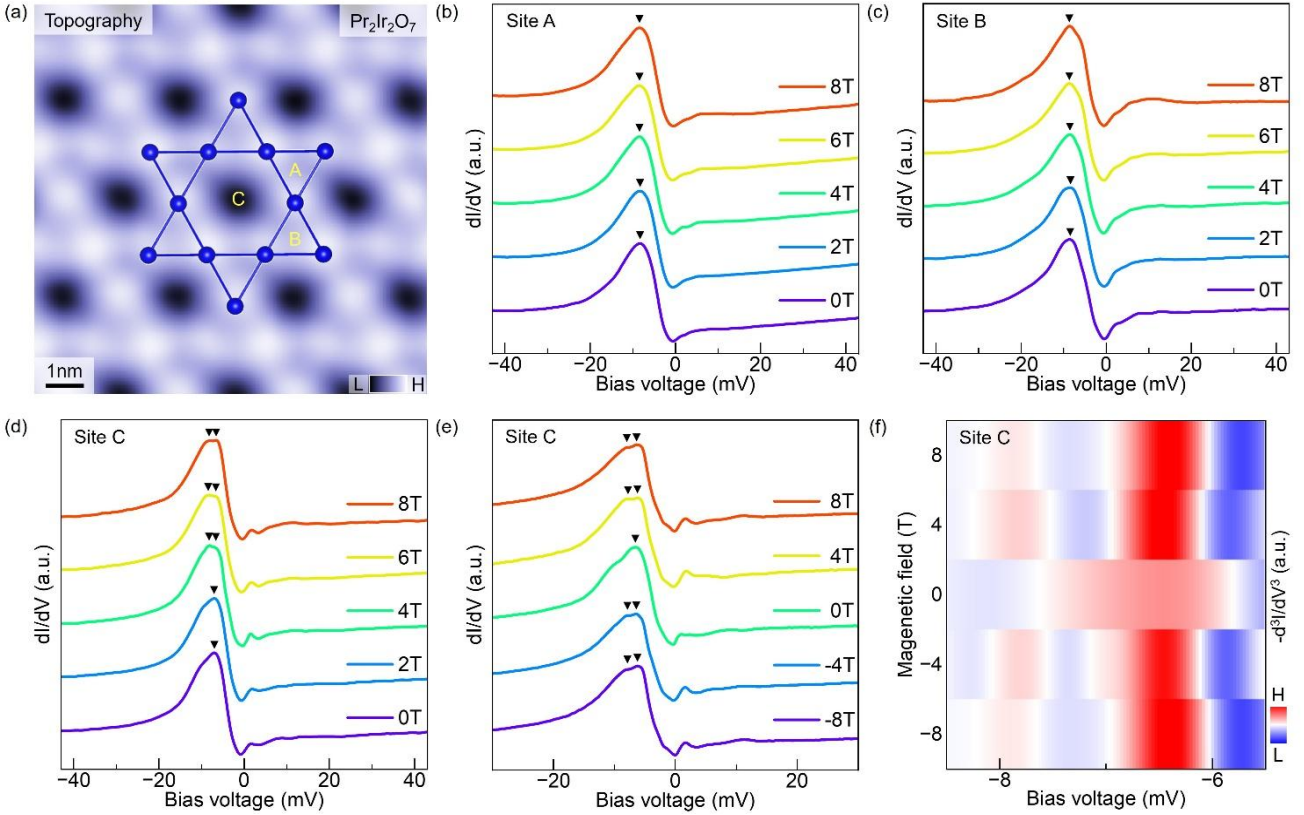

Figure 5 Site dependent Zeeman splitting of the Kondo resonance. (a) Topographic image of the kagome lattice surface indicating measurement sites labeled A, B and C. We use $V = 50$ mV and $I = 500$ pA. (b)(c)(d) Differential conductance spectra taken at site A, B and C under varying magnetic fields up to 8T applied perpendicular to the sample surface. We use $V = 50$ mV, $I = 500$ pA and $V_m = 0.3$ mV. (e) Differential conductance spectra taken at site C measured with reversed magnetic fields. We use $V = 50$ mV, $I = 500$ pA and $V_m = 0.2$ mV. (f) Derivative analysis of magnified differential conductance spectra at site C, highlighting clear Zeeman splitting of the resonance under applied magnetic fields. We use $V = 50$ mV, $I = 500$ pA and $V_m = 0.2$ mV. All data are taken at 0.3 K.

**Acknowledgment**
We acknowledge the discussion with Xianxin Wu, Qianghua Wang and Wen Huang. We acknowledge the support from the National Key R&D Program of China (grant number 2023YFF0718403, 2023YFA1407300), the National Science Foundation of China (grant number 12374060), Guangdong Provincial Quantum Science Strategic Initiative (GDZX2401001). C.S. acknowledges start-up funding from Iowa State University and Ames National Laboratory. H.K. acknowledges support provided by the Deutsche Forschungsgemeinschaft (German Research Foundation), Germany, under Grant No. KU 4080/2-1 (495076551). C.K. acknowledges support by the DFG under Grant No. TRR 360–492547816.

**Data availability**
Data are available from the authors upon reasonable request.